\documentclass[fleqn,10pt]{wlscirep}
\usepackage[utf8]{inputenc}
\usepackage[T1]{fontenc}
\usepackage{newtxtext,newtxmath}% own package
\usepackage{bm}% own package
\title{Trainable Joint Bilateral Filters for Enhanced Prediction Stability in Low-dose CT}

\author[1,*]{Fabian Wagner}
\author[1]{Mareike Thies}
\author[1]{Felix Denzinger}
\author[1]{Mingxuan Gu}
\author[1]{Mayank Patwari}
\author[1]{Stefan Ploner}
\author[1]{Noah Maul}
\author[1]{Laura Pfaff}
\author[1]{Yixing Huang}
\author[1]{Andreas Maier}
\affil[1]{Pattern Recognition Lab, Friedrich-Alexander-Universität Erlangen-Nürnberg, 91058 Erlangen, Germany}

\affil[*]{fabian.wagner@fau.de}

\begin{abstract}
Low-dose computed tomography (CT) denoising algorithms aim to enable reduced patient dose in routine CT acquisitions while maintaining high image quality. Recently, deep learning~(DL)-based methods were introduced, outperforming conventional denoising algorithms on this task due to their high model capacity.
However, for the transition of DL-based denoising to clinical practice, these data-driven approaches must generalize robustly beyond the seen training data.
We, therefore, propose a hybrid denoising approach consisting of a set of trainable joint bilateral filters~(JBFs) combined with a convolutional DL-based denoising network to predict the guidance image. Our proposed denoising pipeline combines the high model capacity enabled by DL-based feature extraction with the reliability of the conventional JBF. The pipeline's ability to generalize is demonstrated by training on abdomen CT scans without metal implants and testing on abdomen scans with metal implants as well as on head CT data.
When embedding two well-established DL-based denoisers (RED-CNN/QAE) in our pipeline, the denoising performance is improved by $10\,\%$/$82\,\%$ (RMSE) and $3\,\%$/$81\,\%$ (PSNR) in regions containing metal and by $6\,\%$/$78\,\%$ (RMSE) and $2\,\%$/$4\,\%$ (PSNR) on head CT data, compared to the respective vanilla model. Concluding, the proposed trainable JBFs limit the error bound of deep neural networks to facilitate the applicability of DL-based denoisers in low-dose CT pipelines.
\end{abstract}
\begin{document}

\flushbottom
\maketitle
\thispagestyle{empty}

\section*{Background}
Minimizing patient dose in computed tomography (CT) is necessary to avoid radiation-related diseases \cite{boone2012radiation}, especially with the number of conducted diagnostic CT scans increasing every year \cite{hess2014trends}. Low-dose CT acquisitions reduce patient dose \cite{wagner2022monte,huang2021semi} but contain higher noise levels in the measured data \cite{barrett1976statistical,maier2015gpu}.
To enhance the image quality of low-dose CT acquisitions, image-based denoising approaches have been proposed, which aim to preserve clinically relevant features compromised with noise. Classical approaches are based on physically motivated conventional filters, considering the inherent properties of the image features \cite{dabov2006image,giraldo2009comparative,tomasi1998bilateral,zhao2019ultra,maier2011three}. Although such filters produce reliable results through a clear algorithmic formulation, their performance is restricted by a limited capability to extract complex features. In addition, conventional filters often require hyperparameters that have to be tuned by hand. Therefore, deep learning~(DL)-based denoising methods gained interest due to their flexibility, strong performance, and data-driven optimization \cite{chen2017low,fan2019quadratic,wu2021low,gu2021adain,li2020sacnn,patwari2020low}. However, deep neural networks usually do not robustly generalize beyond their finite training data distribution, which so far limits clinical applications of DL-based denoising for low-dose CT \cite{antun2020instabilities,hirano2021universal}.\\
Previously, Maier et~al. proved that including physical knowledge in terms of known operators in neural networks reduces the absolute error bound of the model \cite{maier2018precision,maier2019,thies2022calib}. Consequently, different image processing pipelines were proposed, employing physical assumptions about noise characteristics to leverage prediction reliability of DL-based methods in the context of image denoising \cite{patwari2020,wu2018fast}. The joint bilateral filter (JBF) is a conventional denoising filter that allows edge-preserving denoising while considering additional information in terms of a guidance image during its filter operation. Imitating the JBF with a shallow convolutional network led to a reduction of trainable parameters in the JBFnet \cite{patwari2020} and the MJBF architecture \cite{wu2022masked}. Although both architectures are inspired by the JBF operation, they are built from convolutional layers, which raises questions on data integrity and interpretability likewise to other DL methods \cite{wu2018fast}. Other works presented methods to find optimal filter \cite{patwari2022limited} or training \cite{xu2022efficient} hyperparameters by predicting them through external neural networks. However, such approaches do not allow for direct integration into DL models as they can not compute gradients toward those hyperparameters.\\
In our previous work, we presented a trainable bilateral filter with competitive denoising performance that can be included in a differentiable pipeline and optimized in a data-driven fashion \cite{wagner2022ultra}. However, the prediction of bilateral filter layers is solely dependent on three learned spatial parameters and one intensity parameter \cite{tomasi1998bilateral}. Therefore, the bilateral filter operation is conceptually different from the joint bilateral filter algorithm, as JBFs allow considering additional information in terms of a guidance image in their denoising algorithm \cite{petschnigg2004digital}. In this work, we extend our research on bilateral filtering by proposing a fully differentiable, trainable joint bilateral filter that allows denoising using a learned guidance image which broadens its applicability. Our filter layer derives analytical gradients toward the filter input, the image guide, and all filter parameters to achieve differentiability and enable data-driven optimization. Guidance images are estimated using two well-established denoising algorithms: RED-CNN~\cite{chen2017low}, an encoder-decoder architecture achieving competitive performance in recent works~\cite{bera2021noise,huang2021danet}, and Quadratic Autoencoder (QAE)~\cite{fan2019quadratic}, employing quadratic neurons. Our proposed hybrid filter model bridges the gap between deep neural networks' high model capacities and the robustness of conventional denoising filters due to the well-defined, restricted influence of the learned guide.

\subsection*{Contributions}
Our contributions are three-fold. First, we propose a GPU-based, trainable JBF based on an analytical gradient that can be included in any differentiable pipeline. To the best of our knowledge a directly trainable JBF was never presented before. Second, we introduce a hybrid denoising pipeline combining the flexibility of deep neural networks with the robustness of the trainable JBF. Third, we demonstrate the robustness of our model on abdomen CT scans containing metal, with metal not being present in the training data distribution and on out-of-domain head CT scans. Our hybrid JBF-based denoising setting improves the prediction reliability of existing DL-based models with limited computational overhead.

\section*{Methods}

Artificial neural networks are generally trained via gradient descent optimization by minimizing a loss metric $L$ calculated from network predictions to fulfill a desired task \cite{lecun2015deep}. This requires calculating the derivative of the loss $L$ with respect to each trainable model parameter to iteratively update the network during training.
\begin{figure}[htbp]
\centering
\includegraphics[trim=0cm 0.4cm 0cm 0.2cm, width=0.7\linewidth]{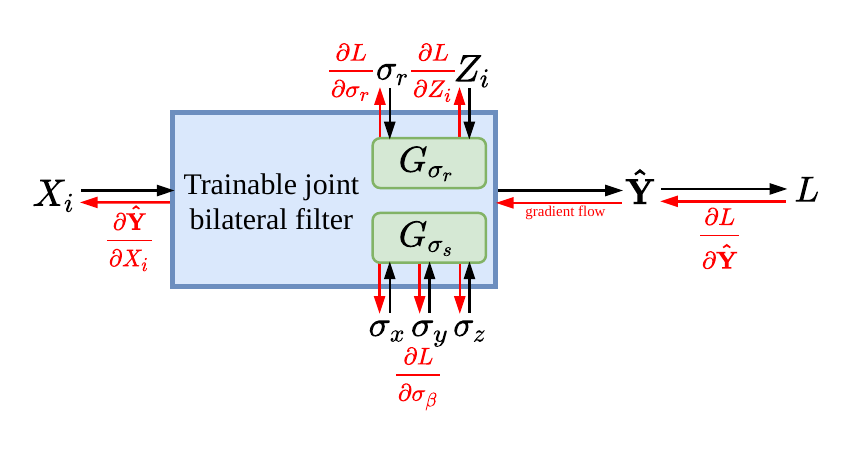}
\caption{Illustration of the proposed trainable joint bilateral filter layer. In the forward pass (black arrows), the input $X_i$ is filtered using parameters $\sigma_\gamma$ $(\gamma \in \{x, y, z, r\})$ and the guidance image $Z_i$ to predict the denoised image $\mathbf{\hat{Y}}$. The model's loss is indicated as $L$. Analytical derivatives are calculated in the backward pass (red arrows) toward filter input, guide, and parameters.} \label{fig:grad_flow}
\end{figure}

\noindent
In this section, the analytical gradient of the proposed trainable JBF layer with respect to filter input, guidance image, and filter parameters is derived as the algorithmic contribution of our work. Figure \ref{fig:grad_flow} illustrates the final working principle of the denoising layer. In the following, bold letters are used to indicate vectors. According to Petschnigg et~al.~\cite{petschnigg2004digital} the JBF operation is defined as
\begin{align}
    \hat{Y}_k = \frac{1}{w_k} \underbrace{\sum_{n \in \mathcal{N}} G_{\sigma_s}(\mathbf{p}_k - \mathbf{p}_n) G_{\sigma_r}(Z_k - Z_n) X_n}_{\substack{=:\,\alpha_k}}
    \label{eq:jbf_def}
\end{align}
and the normalizing factor $w_k$ as
\begin{align}
    w_k := \sum_{n \in \mathcal{N}} G_{\sigma_s}(\mathbf{p}_k - \mathbf{p}_n) G_{\sigma_r}(Z_k - Z_n)\enspace,
\end{align}
with the denoised prediction $\mathbf{\hat{Y}}$ indexed by $k \in \mathbb{N}$, the noisy input image $\mathbf{X}$ in the voxel neighborhood $n \in \mathcal{N}$ around $k$, and a guidance image $\mathbf{Z}$. Guidance images should provide additional information to the filter operation and can be, e.g., additional images paired with the filter input or learned predictions from a neural network as later introduced in this work. The Gaussian intensity range kernel
\begin{align}
    G_{\sigma_r}(c) := \exp \left(-\frac{c^2}{2\sigma_r^2}\right)
\end{align}
is derived from intensity differences on the guidance image $\mathbf{Z}$ and enforces edge sensitivity of the filtering operation. A second, spatial filter kernel $G_{\sigma_s}$ weights voxels according to their spatial distance derived from the positions $\mathbf{p}_k \in \mathbb{N}^d$ and $\mathbf{p}_n \in \mathbb{N}^d$ with $d=3$ for three-dimensional filtering
\begin{align}
    G_{\sigma_s}(\mathbf{c}) = \prod_{s \in \{x,y,z\}} \exp \left(-\frac{c_s^2}{2\sigma_s^2}\right)\enspace.
\end{align}
DL pipelines require gradient calculation of the loss function $L$ with respect to each trainable parameter to enable backpropagation. We can calculate the gradient for our joint bilateral filter layer by using the chain rule
\begin{align}
    \frac{\partial L}{\partial \sigma_\gamma} = \frac{\partial L}{\partial \mathbf{\hat{Y}}} \frac{\partial \mathbf{\hat{Y}}}{\partial \sigma_\gamma}  = \sum_k \frac{\partial L}{\partial \hat{Y}_k} \frac{\partial \hat{Y}_k}{\partial \sigma_\gamma}
\end{align}
with the four kernel widths $\sigma_\gamma$ representing the only trainable weights of the proposed layer when filtering in three dimensions $(\gamma \in \{x, y, z, r\})$. The derivative of the loss function with respect to the filter prediction $\frac{\partial L}{\partial \mathbf{\hat{Y}}}$ is provided by the backpropagation of the loss through differentiable operations applied on the JBF layer output, e.g., subsequent convolutional layers or the loss function itself. The term $\frac{\partial \hat{Y}_k}{\partial \sigma_\gamma}$ can be written using the definition of the joint bilateral filter algorithm from Eq.~\ref{eq:jbf_def} together with the product and chain rule of differentiation
\begin{align}
    \frac{\partial \hat{Y}_k}{\partial \sigma_\gamma} = - w_k^{-2} \alpha_k \frac{\partial w_k}{\partial \sigma_\gamma} + w_k^{-1} \frac{\partial \alpha_k}{\partial \sigma_\gamma}\enspace,
    \label{eq:DerivativeSigI}
\end{align}
the partial derivatives
\begin{align}
    \frac{\partial w_k}{\partial \sigma_\gamma} &= \sum_{n \in \mathcal{N}} \frac{\partial}{\partial \sigma_\gamma} G_{\sigma_s}(\mathbf{p}_k - \mathbf{p}_n) G_{\sigma_r}(Z_k - Z_n)\enspace,\\
    \frac{\partial \alpha_k}{\partial \sigma_\gamma} &= \sum_{n \in \mathcal{N}} X_n \frac{\partial}{\partial \sigma_\gamma} G_{\sigma_s}(\mathbf{p}_k - \mathbf{p}_n) G_{\sigma_r}(Z_k - Z_n)\enspace,
\end{align}
and the Gaussian terms
\begin{align}
    \frac{\partial}{\partial \sigma_\gamma} G_{\sigma}(c) = G_{\sigma}(c) \frac{c^2}{\sigma_\gamma^3}\enspace.
\end{align}
In addition, the derivative of the loss with respect to each input voxel $X_i$ of the joint bilateral filter yields
\begin{align}
    \begin{split}
        \frac{\partial L}{\partial X_i} =& \frac{\partial L}{\partial \mathbf{\hat{Y}}} \frac{\partial \mathbf{\hat{Y}}}{\partial X_i} = \sum_k \frac{\partial L}{\partial \hat{Y}_k} \frac{\partial \hat{Y}_k}{\partial X_i}\\
        =& \sum_k \frac{\partial L}{\partial \hat{Y}_k} w_k^{-1} G_{\sigma_s}(\mathbf{p}_k - \mathbf{p}_i) G_{\sigma_r}(Z_k - Z_i)
    \end{split}
\end{align}
to incorporate the filter as a trainable layer in a differentiable pipeline. The derivative of $L$ with respect to each voxel of the guidance image $Z_i$ can be calculated as
\begin{align}
    \begin{split}
        \frac{\partial L}{\partial Z_i} =& \sum_k \frac{\partial L}{\partial \hat{Y}_k} \frac{\partial \hat{Y}_k}{\partial Z_i} = \sum_k \frac{\partial L}{\partial \hat{Y}_k} \left( - w_k^{-2} \alpha_k \frac{\partial w_k}{\partial Z_i} + w_k^{-1} \frac{\partial \alpha_k}{\partial Z_i} \right)
    \end{split}
\end{align}
where the following two cases must be distinguished, corresponding to arbitrary voxels from the filter neighborhood ($k \neq i$) or the center voxel ($k = i$)

\subsubsection*{Case 1: (\bm{$k \neq i$})}
\begin{align}
    \begin{split}
        \left.\frac{\partial w_k}{\partial Z_i}\right\vert_{k \neq i} =& \,G_{\sigma_s}(\mathbf{p}_k - \mathbf{p}_i) G_{\sigma_r}(Z_k - Z_i) \frac{Z_k - Z_i}{\sigma_r^2}\\
        \left.\frac{\partial \alpha_k}{\partial Z_i}\right\vert_{k \neq i} =& \,G_{\sigma_s}(\mathbf{p}_k - \mathbf{p}_i) G_{\sigma_r}(Z_k - Z_i) \frac{Z_k - Z_i}{\sigma_r^2} X_i
        \label{eq:K_not_I_2}
    \end{split}
\end{align}

\subsubsection*{Case 2: (\bm{$k = i$})}
\begin{align}
    \begin{split}
        \left.\frac{\partial w_k}{\partial Z_i}\right\vert_{k = i} =& \,\sum_{n \in \mathcal{N}} G_{\sigma_s}(\mathbf{p}_i - \mathbf{p}_n) G_{\sigma_r}(Z_i - Z_n) \frac{Z_n - Z_i}{\sigma_r^2}\\
        \left.\frac{\partial \alpha_k}{\partial Z_i}\right\vert_{k = i} =& \,\sum_{n \in \mathcal{N}} G_{\sigma_s}(\mathbf{p}_i - \mathbf{p}_n) G_{\sigma_r}(Z_i - Z_n) \frac{Z_n - Z_i}{\sigma_r^2} X_n\enspace.
        \label{eq:K_is_I_2}
    \end{split}
\end{align}
We calculate the gradients in the backward pass of a fully trainable JBF using the CUDA binding of the \textit{PyTorch} deep learning framework~\cite{paszke2019pytorch} to leverage computational performance. The processing time of one $512 \times 512$ image using $5 \times 5$/$11 \times 11$ pixel kernel windows is around $1.8\,\text{ms}$/$8.0\,\text{ms}$ on the GPU and $69\,\text{ms}$/$350\,\text{ms}$ on the CPU. In comparison, \textit{torch.nn.Conv2d} layers (\textit{PyTorch}) approximately require $0.1\,\text{ms}$/$0.2\,\text{ms}$ (GPU) and $8\,\text{ms}$/$20\,\text{ms}$ (CPU) for processing the single channel image. For both layers gradient calculations have comparable run times as forward passes. All run times were estimated by averaging $50$ repeated forward/backward passes through the respective layers using a \textit{NVIDIA Quadro RTX 4000} GPU. Please note that run times can strongly vary depending on the used hardware.\\
Our filter layer is publicly available at \url{https://github.com/faebstn96/trainable-joint-bilateral-filter-source}. In addition, our code repository contains example scripts and a test script that compares the implementation of the analytical gradient with numerical gradient approximations using the \textit{torch.autograd.gradcheck} function to be sure the filter algorithm is correctly implemented.

\section*{Experimental Setup}

\begin{figure}
\centering
\includegraphics[width=\linewidth]{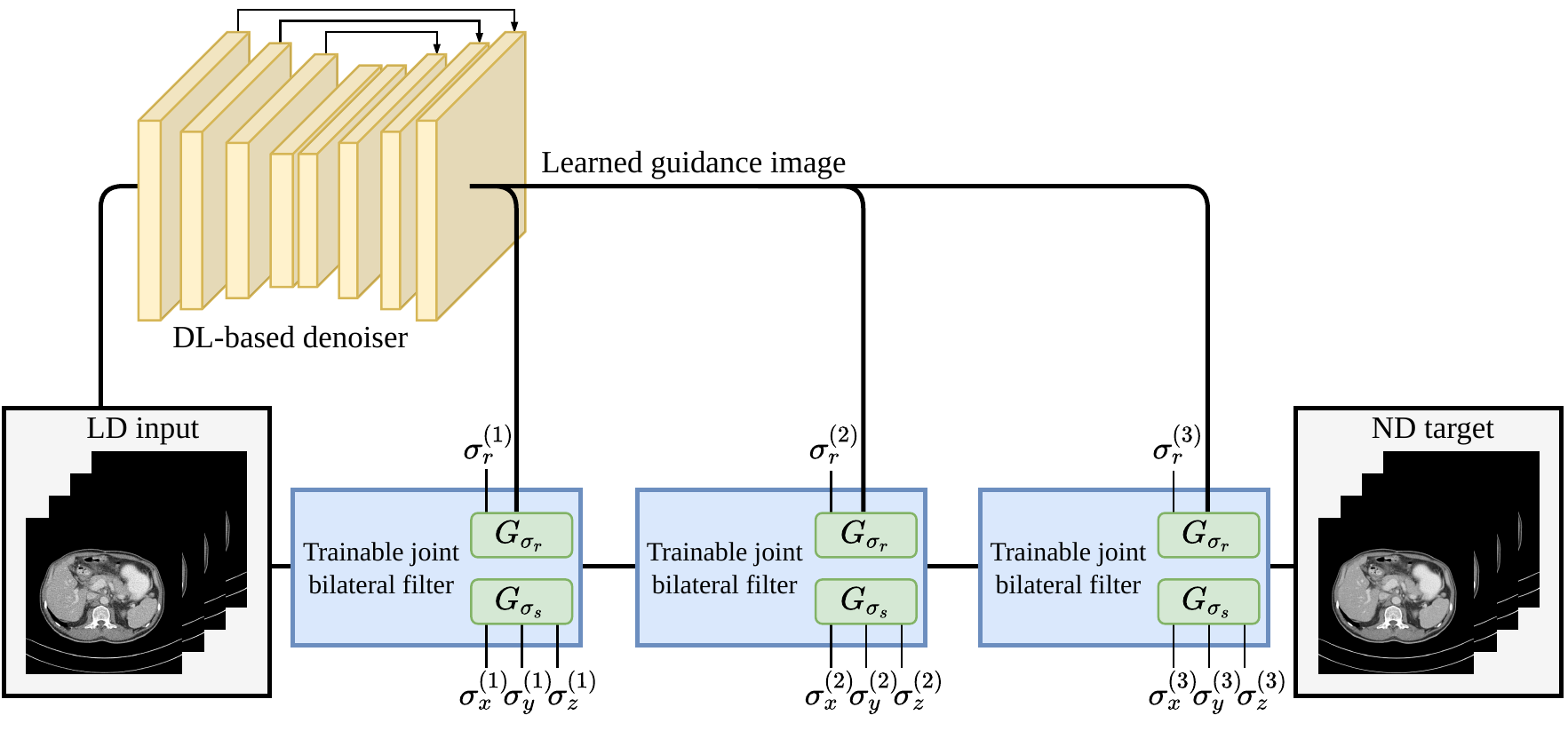}
\caption{The investigated JBF-based denoising pipeline consists of three stacked trainable JBF layers to iteratively remove noise from the low-dose input reconstruction. Pre-trained RED-CNN and QAE models are employed as DL-based denoiser to predict guidance images. The model is trained supervised and tested on CT data from different domains to investigate robustness properties of the pipeline. Indices $(\nu)$ with $\nu \in \lbrace 1, 2, 3\rbrace$ name the individual trainable JBF layers.} \label{fig:setup}
\end{figure}

\subsection*{Denoising Pipeline}
Our denoising pipeline, illustrated in Figure~\ref{fig:setup}, is built on three consecutive trainable JBF layers. The iterative composition of filtering blocks is inspired by the design of the deep convolutional architecture JBFnet \cite{patwari2020} and adds in total twelve independently trainable parameters to the denoising model. The forward pass of each filter layer is calculated as written in Eq.~\ref{eq:jbf_def}. A guidance image is predicted from a deep convolutional network and used to derive the intensity range kernel $G_{\sigma_r}$ in each JBF. Multiple network configurations are presented in the following, investigating the influence of JBF layers on the denoised prediction.

\subsection*{Experiments}
Our experiments are particularly designed to investigate the prediction robustness of hybrid JBF + DL-based denoising models compared to the respective vanilla DL model. We perform experiments with two different well-established low-dose CT denoising architectures predicting the guidance image: RED-CNN \cite{chen2017low} and QAE \cite{fan2019quadratic}. In all our experiments, we train the two reference models independently as described in their works until full convergence of the validation loss after up to $300$ epochs. Subsequently, we place the models in our denoising pipeline and optimize the JBFs for additional $200$ epochs until convergence of the validation loss. Both trained vanilla deep neural networks are used as performance reference. We use the mean squared error loss and two separate Adam optimizers for $\sigma_r$ ($l_r = 1 \cdot 10^{-2}$) and $\sigma_s$ ($l_r = 5 \cdot 10^{-4}$) during training as both sets of parameters define filter kernels that act on independent scales.
\begin{figure}
\centering
\includegraphics[width=0.9\linewidth]{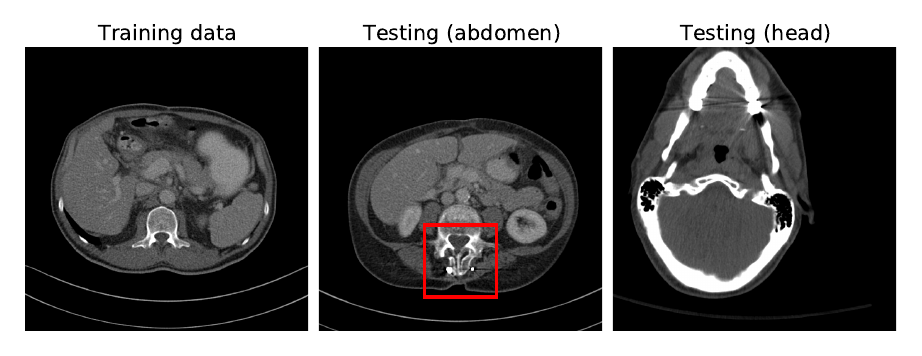}
\caption{Exemplary slices from the training (left, abdomen without metal) and two testing (middle, right) data sets. The abdomen test patients contain regions with metal implants like the one highlighted in red. The reconstruction window is $[-150, 500]\,\text{HU}$.} \label{fig:datasets}
\end{figure}

\subsection*{Data}
All used abdomen and head CT scans are from the public \textit{TCIA Low Dose CT Image and Projection} data set (Version 4) \cite{moen2021low}, containing paired low-dose ($25\,\%$ dose) and high-dose CT volumes. The goal of our experiment is to quantitatively evaluate the robustness of the introduced denoising models and compare them with the vanilla DL-based denoising models RED-CNN and QAE. Therefore, we manually split the abdomen data into two domains. First, patients without metal pieces and second reconstructions containing pieces of metal like implants or catheters that appear as bright regions due to their strong x-ray absorption. Only data from the first domain not containing metal is used for training (21 scans) and validation (two scans). Subsequently, we test our models on the previously unseen metal domain scans (24 scans) to evaluate how the different architectures can handle examples that are insufficiently represented by the training data domain. As the metal pieces are usually located in small sub-volumes of the reconstructions, we additionally define 17 three-dimensional regions of interest (ROIs) that are evaluated separately to get more expressive results on the sensitivity to out-of-domain features. The coordinates of all 17 ROIs are provided in the supplementary material together with exemplary abdomen slices containing the respective ROIs to facilitate reproducibility. Additionally, we test our models on data from a separate domain, namely head CT scans (20 scans), to investigate prediction robustness on a different anatomy. Figure~\ref{fig:datasets} shows example slices from the training and testing data sets with a highlighted abdomen ROI containing metal parts. Note that all scans are directly taken from the public data set without further modification such that they well represent clinical routine head and abdomen CT acquisitions of patients with and without metal implants \cite{moen2021low}.

\section*{Results}

\subsection*{Quantitative Results}
\begin{table}[htbp]
\centering
\begin{tabular}{lcccccc}
& \multicolumn{3}{c }{\textbf{Full metal data}} & \multicolumn{3}{c}{\textbf{Metal ROIs}} \\
\cmidrule(lr){2-4} \cmidrule(lr){5-7}
&  \textbf{RMSE} $[\text{HU}]$ $\downarrow$ &  \textbf{PSNR} $\uparrow$ & \textbf{SSIM} $\uparrow$  &  \textbf{RMSE} $[\text{HU}]$ $\downarrow$ & \textbf{PSNR}  $\uparrow$ &  \textbf{SSIM} $\uparrow$ \\
\hline
Low-dose CT
& $18.6 \pm 6.1$ & $41.51 \pm 2.13$ &  $0.9512 \pm 0.031$ & $26.7 \pm 15.9$ &  $38.98 \pm 4.28$ & $0.9594 \pm 0.030$ \\
\hline
RED-CNN \cite{chen2017low}
& \bm{$12.0 \pm 3.9$} & \bm{$45.38 \pm 1.93$} & \bm{$0.9816 \pm 0.012$} & $24.6 \pm 13.1$ &  $39.68 \pm 3.75$ & $0.9788 \pm 0.015$ \\
\textbf{RED$+$JBFs}
& $12.7 \pm 4.5$ & $45.01 \pm 2.02$ & $0.9797 \pm 0.016$ & \bm{$22.1 \pm 14.2$} &  \bm{$40.82 \pm 4.27$} & \bm{$0.9797 \pm 0.013$} \\
\hline
QAE \cite{fan2019quadratic}
& $44.7 \pm 36.1$ & $43.17 \pm 4.49$ & $0.9811 \pm 0.014$ & $837.8 \pm 478.3$ &  $14.88 \pm 10.92$ & $0.6967 \pm 0.232$ \\
\textbf{QAE$+$JBFs}
& \bm{$14.5 \pm 5.4$} & \bm{$44.61 \pm 2.23$} & \bm{$0.9827 \pm 0.010$} & \bm{$150.9 \pm 87.5$} &  \bm{$26.99 \pm 7.19$} & \bm{$0.9075 \pm 0.062$} \\
\hline
\end{tabular}
\caption{Quantitative denoising results on the full abdomen scans containing metal implants as well as on 17 ROIs that contain metal parts. The metrics are averaged over all test patients or all test ROIs and provided as $\text{mean} \pm \text{std}$. The respectively better performing network modification is highlighted in bold.}\label{tab:quant_results}
\end{table}
\begin{figure}
\centering
\includegraphics[width=\linewidth]{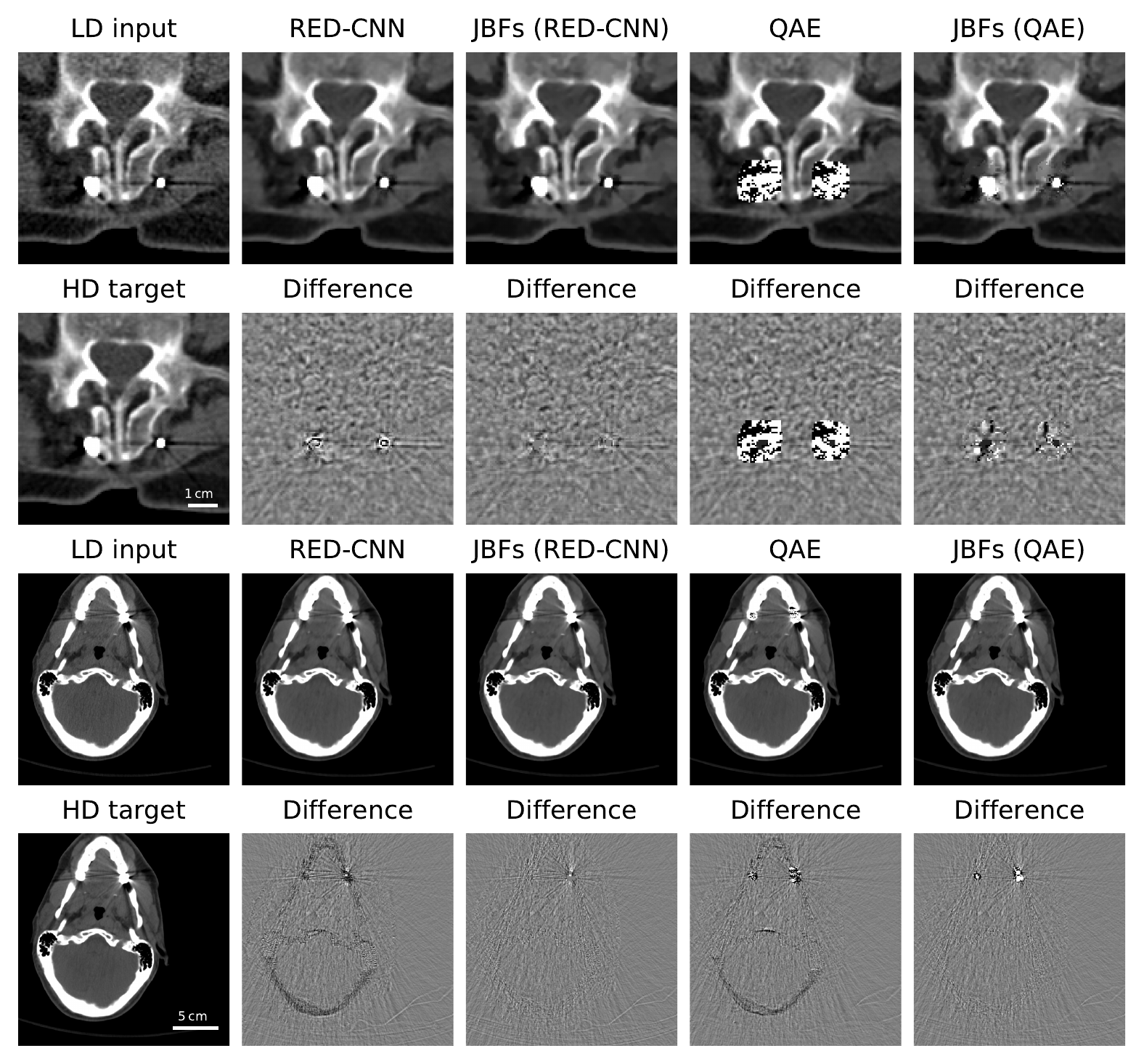}
\caption{Qualitative denoising results on the ROI highlighted in Fig.~\ref{fig:datasets} and on a head CT slice. Difference images are calculated between model prediction and high-dose (HD) target and are shown in the window $[-50, 50]\,\text{HU}$ for abdomen data and $[-100, 100]\,\text{HU}$ for head data. The reconstruction window is $[-150, 500]\,\text{HU}$. Our hybrid models visually outperform the respective vanilla deep models.
} \label{fig:vis_results}
\end{figure}
\noindent
We present quantitative denoising results on the entire abdomen test data set and only on the abdomen ROIs containing metal pieces in Table~\ref{tab:quant_results}. Performance metrics for the investigated out-of-domain head CT data set are listed in Table~\ref{tab:quant_results_head}. The three established image quality metrics root-mean-square error (RMSE), peak signal-to-noise ratio (PSNR), and structural similarity (SSIM) \cite{wang2004image} are calculated to compare model prediction with their respective high-dose target reconstruction. RMSE and PSNR particularly assess deviations from the target image intensities, whereas SSIM aims to imitate human perception to compare image content. We found that all performance differences between vanilla and respective JBF-based model in Table~\ref{tab:quant_results} are significant based on a Wilcoxon signed-rank test \cite{wilcoxon1992individual} on a p-value $p < 0.005$. The Wilcoxon signed-rank test is particularly suited to test the paired model predictions at hand without presuming an underlying statistical model.\\
Whereas the hybrid JBF layer-based pipelines perform comparably to the vanilla deep denoising models over the entire abdomen test data, an explicit performance improvement is recognized on the 17 abdomen ROIs as well as on the head CT scans on all three investigated image quality metrics. Both JBF-based pipelines decrease the RMSE by $10\,\%$/$82\,\%$ and improve the PSNR and SSIM by $3\,\%$/$81\,\%$ and $0.1\,\%$/$30\,\%$ around the out-of-training-domain metal features compared to the vanilla RED-CNN and QAE respectively. The denoising performance on the head CT data is improved by $6\,\%$/$78\,\%$ (RMSE), $2\,\%$/$4\,\%$ (PSNR), and $0.1\,\%$/$0.1\,\%$ (SSIM).

\begin{table}[htbp]
\centering
\begin{tabular} { l c c c }
%\hline
& \multicolumn{3}{c}{\textbf{Head CT data}} \\
\cmidrule(lr){2-4} 
&  \textbf{RMSE} $[\text{HU}]$ $\downarrow$ &  \textbf{PSNR} $\uparrow$ & \textbf{SSIM} $\uparrow$ \\
\hline
Low-dose CT
& $5.3 \pm 1.1$ & $38.36 \pm 1.58$ &  $0.9338 \pm 0.017$ \\
\hline
RED-CNN \cite{chen2017low}
& $5.1 \pm 0.6$ & $38.42 \pm 0.88$ &  $0.9646 \pm 0.006$ \\
\textbf{RED$+$JBFs}
& \bm{$4.8 \pm 0.6$} & \bm{$39.01 \pm 0.98$} &  \bm{$0.9662 \pm 0.006$} \\
\hline
QAE \cite{fan2019quadratic}
& $44.8 \pm 40.9$ & $37.30 \pm 2.58$ &  $0.9679 \pm 0.007$ \\
\textbf{QAE$+$JBFs}
& \bm{$9.8 \pm 6.4$} & \bm{$38.97 \pm 1.77$} &  \bm{$0.9695 \pm 0.007$} \\
\hline
\end{tabular}
\caption{Quantitative denoising results on the full head CT data set. The metrics are averaged over all test patients and provided as $\text{mean} \pm \text{std}$. The respectively better performing network modification is highlighted in bold.}\label{tab:quant_results_head}
\end{table}

\subsection*{Qualitative Results}
Visual results on one ROI and a head CT slice are displayed in Figure~\ref{fig:vis_results}. Provided difference images between model prediction and high-dose target particularly highlight disturbed features and erroneous predictions. Intensity distortions in close proximity to metal implants and in the skull region can be recognized for the RED-CNN, which get almost entirely removed using the RED-CNN prediction as guidance image in a JBF-based setting. Here, in particular the intensity shifts visible as a shadows of the skull in the difference images of the vanilla model prediction are fully restored by the proposed hybrid JBF-based model.\\
The QAE predicts strong artifacts that are visible in the abdomen intensity images and difference images surrounding metal implants. Using such predictions as an image guide in a JBF-based pipeline produces results that visually look much closer to the high-dose target where features like the shape of metal pieces or the adjacent anatomy are visible. Further, intensity distortions in QAE predictions on the head CT data set are removed using the combined QAE+JBFs filtering approach. Only regions around the dental crowns with heavy metal reconstruction artifacts remain disturbed.

\section*{Discussion} Although one could simply add abdomen scans containing metal pieces or head CT data to the training data set to improve denoising performance, our experiment is particularly designed to evaluate and quantify robustness to real CT data that is underrepresented in the training data. Our experiment, therefore, mimics the present clinical scenario where a model is only trained on a limited number of studies but must also handle differing anatomies and scanning parameters. The denoising performance of a JBF depends on an optimal intensity range kernel $G_{\sigma_r}$ to avoid blurring edges. Here, the proposed pipeline can benefit from the guidance image that is predicted by a deep model that is capable of employing global image features to facilitate extracting sharp edges needed for the filter kernel computation. In case of prediction failures like in regions around metal implants or at the skull, the intensity range kernel contribution is either over- or underestimated. This results in over- or under-smoothing of the respective image region but is always based on the local content of the input image. Therefore, the intensity range kernel design of Gaussian shape prevents the output from large prediction errors by design.\\
In our conducted experiments, pre-trained denoising networks predict the guidance images that are input to the JBF layers. We performed additional experiments, training the JBFs together with the denoising networks in a combined end-to-end setting. Although this setting enhanced performance within the training data domain, we did not recognize explicit performance improvements in terms of robustness on the investigated out-of-training-domain data sets. Eventually, we did not design our experiments to answer the question how a guidance image that is optimized for JBFs is handled in the training data domain but we particularly want to investigate how JBFs handle the displayed artifacts predicted by the denoising networks as the primary goal of our study.\\
JBF-based pipelines almost entirely prevent the predictions from artifacts introduced by the DL-based models but the combined QAE+JBFs predictions still contain some slight distortions around the spine metal implant in Figure \ref{fig:vis_results}. These results visualize that the JBF, although enforcing proximity to the noisy input, is dependent on a reasonable guidance image. However, the shown artifacts introduced by the QAE network can be regarded as worst-case in a clinical pipeline and are still satisfactorily handled by the JBFs compared to the initial QAE prediction.\\
DL frameworks like \textit{PyTorch}~\cite{paszke2019pytorch} allow an automatic calculation of gradients in their operators. Therefore, one could think of implementing a JBF directly from \textit{PyTorch} tensors instead of using analytical gradients to make its parameters trainable. Although this is possible, training such a filter would require expensive Python loops over the training batches and kernel windows which would accumulate huge computational graphs for the gradient calculation. In practice, training such a model with reasonable image and batch sizes, therefore, is infeasible in terms of computational time and GPU memory. The analytical filter derivative presented in this work greatly simplifies the required computations to enable data-driven optimization and limit the computational overhead through adding JBF layers as shown by comparing run times with convolutional layers. Eventually, we believe that our open-source filter layer can be useful in further hybrid applications as a known denoising operator that can be optimized in a data-driven manner.

\section*{Conclusion}
In this work, we presented a trainable JBF layer that can be incorporated into any deep model. We propose a hybrid denosing pipeline using these JBF layers and pre-trained deep denoising neural networks. The latter can produce faulty predictions when tested on data that is insufficiently represented in the training domain. In our experiments, we show that JBFs prevent DL-based models from severe prediction failures although the JBFs make use of distorted guidance images predicted from the neural networks. These results are explained by the clear algorithmic design of the JBF that limits the influence of the guidance image to the contribution of the intensity range filter kernel. We think that JBF layers can combine the flexibility of deep neural networks with the prediction reliability of conventional methods to leverage the power of deep models in clinical low-dose CT applications.

\section*{Data availability}
The data sets analysed during the current study are publicly available in the \textit{TCIA Low Dose CT Image and Projection Data} repository (Version 4) \cite{moen2021low}, \url{https://doi.org/10.7937/9NPB-2637}. Coordinates and exemplary slices of all analyzed abdomen ROIs are included in the supplementary material.

\section*{Code availability}
The implementation of our open-source CUDA-accelerated trainable bilateral filter layer (\textit{PyTorch}) together with example scripts and tests is publicly available at \url{https://github.com/faebstn96/trainable-joint-bilateral-filter-source}.

\bibliography{sample}

\section*{Author contributions statement}

F.W. conceived and conducted the experiments and the algorithm derivation. M.T., F.D., and M.G. contributed on the filter algorithm and experimental design. M.P. provided network implementations. S.P. assisted with the CUDA kernels. N.M., L.P., and Y.H. provided valuable technical feedback during development. All authors reviewed the manuscript.

\section*{Acknowledgements}

The research leading to these results has received funding from the European Research Council (ERC) under the European Unions Horizon 2020 research and innovation program (ERC Grant No. 810316). Further, we thank the NVIDIA Corporation for their GPU donation through the NVIDIA Hardware Grant Program.

\section*{Competing interests}

The authors declare no competing interests.

\end{document}

\flushbottom
\maketitle

\begin{table}[htbp]
\centering
\begin{tabular}{lllllll}
\hline
\textbf{Scan ID} & \textbf{x start} & \textbf{x end} & \textbf{y start} & \textbf{y end} & \textbf{z start} & \textbf{z end} \\
\hline
L006 &  $190$ & $240$ & $170$ & $210$ &	$204$ & $214$ \\
L033 &  $150$ & $380$ & $95$  & $130$ &	$0$   & $11$  \\
L049 &  $205$ & $320$ & $275$ & $380$ & $6$   & $9$   \\
L058 &  $450$ & $500$ & $230$ & $300$ &	$205$ & $209$ \\
L075 &  $200$ & $315$ & $160$ & $250$ &	$16$  & $28$  \\
L077 &  $215$ & $300$ & $455$ & $480$ &	$7$   & $9$   \\
L107 &  $230$ & $340$ & $430$ & $460$ &	$74$  & $82$  \\
L114 &  $460$ & $500$ & $190$ & $250$ &	$10$  & $15$  \\
L134 &  $70$  & $170$ & $220$ & $360$ &	$6$   & $7$   \\
L148 &  $35$  & $240$ & $135$ & $260$ &	$0$   & $38$  \\
L160 &  $215$ & $285$ & $280$ & $340$ &	$99$  & $114$ \\
L170 &  $175$ & $330$ & $205$ & $340$ &	$37$  & $115$ \\
L178 &  $105$ & $185$ & $245$ & $310$ &	$0$   & $7$   \\
L187 &  $210$ & $300$ & $375$ & $420$ &	$33$  & $45$  \\
L193 &  $200$ & $305$ & $175$ & $250$ &	$161$ & $168$ \\
L203 &  $200$ & $415$ & $240$ & $325$ &	$11$  & $55$  \\
L232 &  $210$ & $270$ & $270$ & $310$ &	$151$ & $159$ \\
\hline
\end{tabular}
\caption{Coordinates of the 17 investigated abdomen ROIs in scans of the \textit{TCIA Low Dose CT Image and Projection} data set (Version 4) \cite{moen2021low}. ROIs are defined by their start and end pixel index (starting from $0$) in all three dimensions. Figure \ref{fig:roi_overview} illustrates exemplary CT slices that contain each ROI.}\label{tab:roi_coordinates}
\end{table}

\begin{figure}
\centering
\includegraphics[width=\linewidth]{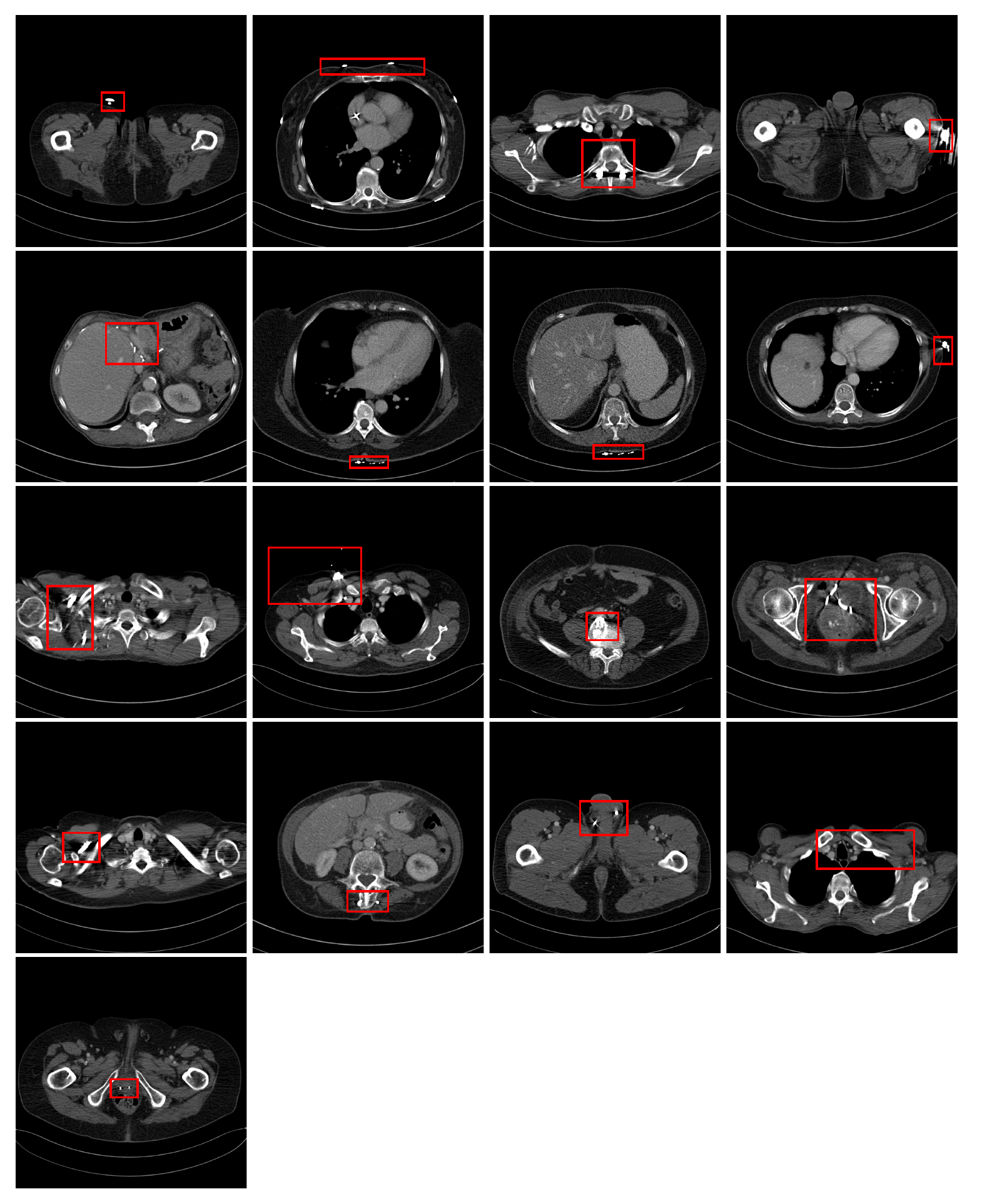}
\caption{Exemplary low-dose input slices with all 17 selected 3D abdomen ROIs (highlighted in red) that contain metal parts. The patients from Table \ref{tab:roi_coordinates} are plotted from left to right and from top to bottom. Note that all ROIs are chosen around three-dimensional objects such that metal pieces are not always found in the center of some of the displayed ROIs. The reconstruction window is $[-150, 500]\,\text{HU}$.} \label{fig:roi_overview}
\end{figure}

\newpage
\bibliography{sample_supplementary}